\documentclass{article}

\usepackage[preprint]{neurips_2024}

\usepackage[utf8]{inputenc} 
\usepackage[T1]{fontenc}    
\usepackage{hyperref}       
\usepackage{url}            
\usepackage{booktabs}       
\usepackage{amsfonts}       
\usepackage{nicefrac}       
\usepackage{microtype}      
\usepackage{xcolor}         
\usepackage{amsmath}
\usepackage{graphicx}       
\usepackage{enumitem}
\usepackage{pifont}

\title{Next Tokens Denoising for Speech Synthesis}

\author{%
    \textbf{Yanqing Liu$^{*}$} \quad
    \textbf{Ruiqing Xue$^{*}$} \quad
    \textbf{Chong Zhang} \quad
    \textbf{Yufei Liu} \quad
    \textbf{Gang Wang} \\
    \textbf{Bohan Li} \quad
    \textbf{Yao Qian} \quad
    \textbf{Lei He} \quad
    \textbf{Shujie Liu} \quad
    \textbf{Sheng Zhao} \\
    \\ Microsoft
}

\begin{document}

\maketitle
\def\thefootnote{*}\footnotetext[1]{These authors contributed equally to this work. Corresponding author: Yanqing Liu, yanqliu@microsoft.com}\def\thefootnote{\arabic{footnote}}

\begin{abstract}

While diffusion and autoregressive (AR) models have significantly advanced generative modeling, they each present distinct limitations. AR models, which rely on causal attention, cannot exploit future context and suffer from slow generation speeds. Conversely, diffusion models struggle with key-value (KV) caching. To overcome these challenges, we introduce Dragon-FM, a novel text-to-speech (TTS) design that unifies AR and flow-matching. This model processes 48 kHz audio codec tokens in chunks at a compact rate of 12.5 tokens per second. This design enables AR modeling across chunks, ensuring global coherence, while parallel flow-matching within chunks facilitates fast iterative denoising. Thus, the model leverages KV-cache across chunks and utilizes bidirectional context within each chunk. Furthermore, it bridges continuous and discrete feature modeling, demonstrating that continuous AR flow-matching can predict discrete tokens with finite scalar quantizers. This efficient codec and fast chunk-autoregressive architecture also make the model highly effective for generating long-form content, such as podcasts. Experiments\footnote{See \url{https://cognitivespeech.github.io/dragon-fm} for demos of our work} on podcast datasets demonstrate its capability to efficiently generate high-quality zero-shot podcasts.


\end{abstract}

\begin{figure}[!htbp]
\centering
\includegraphics[width=1.0\textwidth]{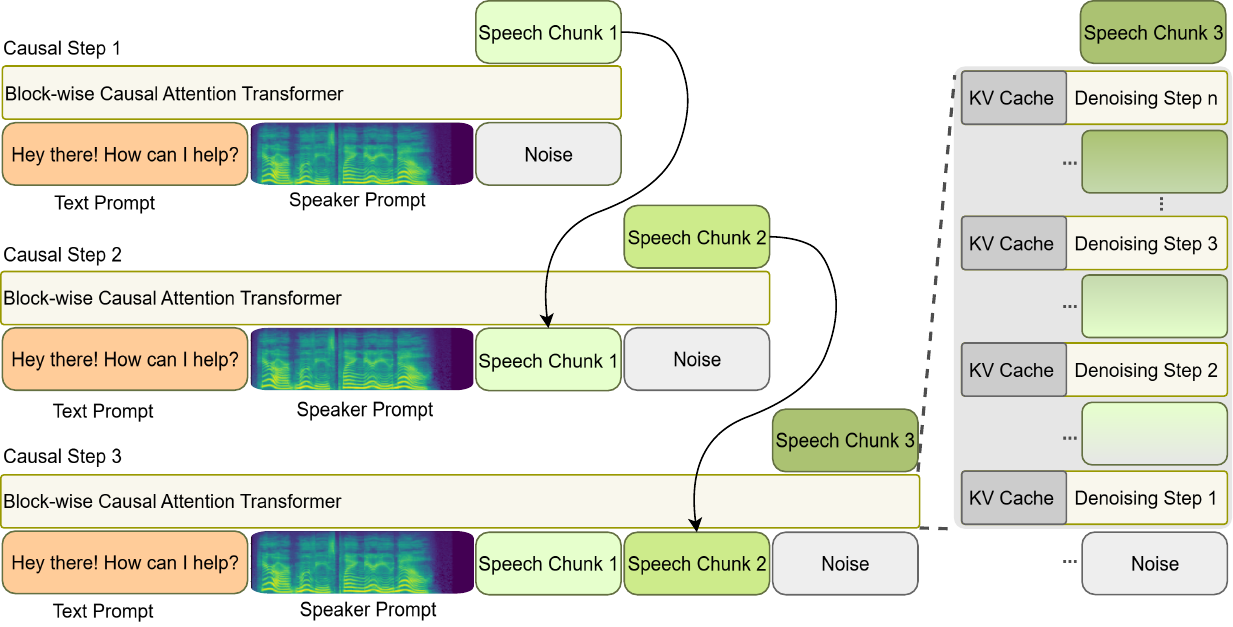}
\caption{Proposed Architecture. Dragon-FM autoregressively predicts chunk-level speech tokens, with text and speech prompt. Within each block, flow-matching enables parallel prediction.}
\label{fig:overview}
\end{figure}
\section{Introduction}

Generative modeling is characterized by a diverse and increasingly fragmented landscape of architectures and feature modeling approaches. Among these, autoregressive (AR) and diffusion models \citep{radford2018improving,lipman2022flow,furuta2024improving,wang2023neural, ramesh2021zero} are leading paradigms across audio, image, and video generation. While both share iterative architectures, they exhibit distinct limitations, fundamentally differing in their iteration strategies, attention mechanisms, and token granularity. Diffusion models achieve higher inference throughput by generating multiple tokens in parallel through iterative denoising with bidirectional attention. Conversely, AR models predict tokens sequentially using causal attention, efficiently leveraging key-value (KV) caching for lower latency—a capability diffusion models currently lack. Regarding generation quality, diffusion models excel in continuous feature modeling, capturing detailed reconstructions with full-context attention but often struggle with discrete token modeling. In contrast, AR models, widely adopted in text generation (e.g., GPT series) \citep{brown2020language}, offer efficient discrete sampling and diverse outputs. However, their reliance on causal attention restricts access to future token information, potentially limiting performance in tasks requiring global context, such as reasoning or multimodal synthesis.

Features used in diffusion and autoregressive (AR) models for various modalities are commonly categorized as discrete or continuous. Discrete tokens, whether from byte-pair encoding (BPE) for text or variational autoencoders (VAEs) with quantizers for audio, images, and video, offer advantages in compression and semantic richness, making them well-suited for next-token prediction in language models. A fundamental challenge lies in balancing training efficiency with compression quality. Higher token rates improve reconstruction but complicate long-sequence training for large language models (LLMs), while shorter sequences risk compromising reconstruction quality, potentially reducing intelligibility and increasing word error rates (WER) in speech. Conversely, continuous features convey richer acoustic and visual details but are inherently more difficult to model. While discrete AR language models perform exceptionally well in text generation due to their manageable BPE-encoded sequence lengths, discrete AR models for speech or images frequently struggle with the extensive token sequences required, often resulting in degraded reconstruction quality. In contrast, diffusion models excel at continuous text-to-image or video generation by effectively leveraging full-context modeling. Yet, both autoregressive and diffusion approaches face a compounding computational burden due to their iterative nature.

To address the aforementioned limitations, we introduce a novel framework and validate its effectiveness through text-to-speech (TTS) applications. Recent advancements in TTS leverage discrete autoregressive (AR) codec language models and continuous flow-matching approaches to balance prosody modeling with fine-grained audio quality. For example, VALL-E \citep{wang2023neural} employs two separately trained discrete language models: one autoregressively predicts residual codes of the first codec layer, while the other non-autoregressively predicts subsequent layers, leading to high latency. Tortoise \citep{betker2023better} attempts to reduce latency by using a shorter discrete token sequence via a mel quantizer and replacing its second model with a diffusion-based approach, which improves real-time factor (RTF) compared to VALL-E but struggles with separate model optimization. E2 \citep{eskimez2024e2}, a single-stage flow-matching model, directly predicts mel spectrograms instead of codec tokens, achieving high RTF without KV-cache but with reduced prosodic variation compared to AR models due to its fixed duration. Other efforts in TTS focus on improving codec token rates, developing advanced quantizers, and designing hybrid AR-diffusion systems, which often require extensive fine-tuning or complex architectural engineering. Quantizing speech into discrete sequences is inherently more challenging than text byte-pair encoding (BPE) due to speech’s rich prosody, timbre, and accent variations, necessitating specialized approaches like hierarchical decoders for layered codecs and cascaded speech upsampling with diverse AR and diffusion combinations. \citep{jia2025ditar} compresses continuous token patches using an aggregation encoder and employs a language model for inter-patch prediction, while a diffusion transformer handles intra-patch prediction with historical token patches. However, this aggregation and historical patch setting may affect sequence modeling quality due to compression artifacts. \citep{liu2024autoregressive} proposes ARDiT, a decoder-only diffusion transformer model that autoregressively generates continuous speech features. However, ARDiT needs a given total duration of target speech during inference and incurs larger latency due to longer speech sequences. Table \ref{tab:table1} summarizes the comparison of existing AR and diffusion-based TTS models.

\begin{table}[h]
\centering
\setlength{\tabcolsep}{8pt} 
\renewcommand{\arraystretch}{1.5} 
\begin{tabular}{ccccccc} 
\toprule
\textbf{Model} & \textbf{AR} & \textbf{Diffusion} & \textbf{E2E} & \textbf{Low Frame Rate} & \textbf{Low Latency} & \textbf{Low RTF} \\ 
\midrule
VALL-E & \textcolor{green}{\ding{51}} & \textcolor{red}{$\times$} & \textcolor{red}{$\times$} & \textcolor{red}{$\times$}  & \textcolor{red}{$\times$} & \textcolor{red}{$\times$}   \\
TorToise & \textcolor{green}{\ding{51}} & \textcolor{green}{\ding{51}} & \textcolor{red}{$\times$}    & \textcolor{red}{$\times$} & \textcolor{red}{$\times$} & \textcolor{red}{$\times$}   \\
E2 & \textcolor{red}{$\times$} & \textcolor{green}{\ding{51}} & \textcolor{green}{\ding{51}} & \textcolor{red}{$\times$}  & \textcolor{red}{$\times$} & \textcolor{green}{\ding{51}}  \\ 
ARDiT & \textcolor{green}{\ding{51}} & \textcolor{green}{\ding{51}} & \textcolor{green}{\ding{51}} & \textcolor{red}{$\times$}  & \textcolor{green}{\ding{51}}  & \textcolor{green}{\ding{51}}   \\ 
Dragon-FM & \textcolor{green}{\ding{51}} & \textcolor{green}{\ding{51}} & \textcolor{green}{\ding{51}}   & \textcolor{green}{\ding{51}}  & \textcolor{green}{\ding{51}}  & \textcolor{green}{\ding{51}}    \\
\bottomrule
\end{tabular}
\vspace{10pt}
\caption{Comparison between the proposed model and state-of-the-art AR and diffusion-based TTS systems. Here, AR refers to autoregressive modeling, and diffusion indicates diffusion-based models. End-to-end (E2E) denotes joint training of the acoustic model. Low latency indicates small first-byte latency, approaching streaming systems, while low real-time factor (RTF) reflects strong parallel processing capability. We define a low frame rate as fewer than 15 tokens per second.}
\label{tab:table1}
\end{table}

In this work, we propose a novel generative modeling and text-to-speech (TTS) framework, Dragon-FM (flow matching), designed to overcome the limitations of individually optimized autoregressive (AR) and diffusion models. Our framework incorporates a high-quality 48kHz codec that compresses audio into a discrete token sequence at 12.5 tokens per second using a finite scalar quantizer \citep{mentzer2023finite, parker2024scaling}, ensuring high reconstruction quality and manageable sequence lengths. A key innovation of the proposed model lies in its integration of AR next-token prediction with parallel flow-matching denoising within the same token chunk. This significantly reduces iterative steps while simultaneously preserving AR’s crucial key-value (KV) caching efficiency and leveraging flow-matching’s bidirectional context attention to boost scalability and sampling diversity. Recognizing analogous challenges in text modeling, where long-sequence large language models (LLMs) hinder real-time applications and diffusion models improve performance at an efficiency cost, our acoustic model directly addresses these. It predicts discrete codec tokens via a continuous flow-matching approach, processing 2-second blocks (25 audio tokens) generated simultaneously through flow-matching, with consecutive blocks produced autoregressively, similar to language models. We term this approach 'next-token denoising,' a new paradigm that unifies continuous and discrete features of LLMs and diffusion models, contrasting with traditional discrete token-based next-token prediction. Extensive experiments validate the proposed codec and acoustic model design, powerfully demonstrating the impact of its optimized speech token rates on both latency and performance.

Our contributions are summarized as follows:
\begin{itemize}
\item We developed a high-quality 48 kHz audio codec using finite scalar quantizers, producing compact 12.5 discrete speech tokens per second, which facilitates efficient AR and DiT-based acoustic modeling.
\item We introduced a novel acoustic model unifying AR and flow-matching, predicting audio token chunks autoregressively while applying few-step, parallel flow-matching within each chunk to significantly reduce iteration steps.
\item We proposed a unified approach for discrete and continuous feature prediction via a finite scalar quantization feature normalizer with autoregressive flow-matching, enhancing sampling variability and advancing multi-modal generative modeling.
\end{itemize}

\section{Related work}

\subsection{Text-to-Speech}

Text-to-speech (TTS) systems have significantly enhanced tone, prosody, fidelity, and intelligibility, shifting from continuous feature modeling \citep{li2018close,li2019neural,li2020robutrans,li2020moboaligner,chen2021adaspeech,liu2021delightfultts,zhang2022mixed,tan2024naturalspeech,kim2021conditional,wang2025streammel,sun2025zero, yuan2024continuous,eskimez2024e2,meng2024autoregressive,liu2022delightfultts}, such as mel spectrograms, to hybrid approaches that integrate discrete and continuous representations \citep{wang2023neural,chen2024vall,xue2023foundationtts,zhang2023speak,shen2023naturalspeech,leng2023prompttts,kanda2024making,ju2024naturalspeech,zhang2024boosting,han2024vall,eskimez2024e2,li2024investigating,yang2025pseudo,wang2025felle}. Continuous feature approaches, utilizing autoregressive (AR) or non-autoregressive (NAR) transformers and diffusion models, excel at capturing fine-grained acoustic details. Discrete token-based systems, typically powered by AR language models, enable scalable and robust zero-shot TTS capabilities. Continuous speech features, derived from signal processing techniques like mel-spectrogram extraction or models such as VQ-VAE codecs, capture high-dimensional compressed representations from raw audio. In contrast, discrete speech tokens are quantized into finite token IDs using residual variational autoencoders (VAEs). Continuous features are modeled with transformer-based AR or NAR architectures and diffusion models optimized via regression loss, whereas discrete tokens are predicted using AR language models. Recently, zero-shot TTS has gained prominence, with systems like E2 \citep{eskimez2024e2} (leveraging continuous features) and VALL-E \citep{chen2024vall} (utilizing discrete tokens) achieving superior zero-shot quality through large-scale models and extensive training data.

\subsection{Flow Matching for Generative Modeling}

Flow-matching \citep{lipman2022flow,tong2023improving,liu2022flow} learns a time-dependent vector field to transform probability distributions, providing an efficient method for modeling ordinary differential equations (ODEs) through straightforward vector-field regression loss. By incorporating optimal transport principles \citep{khrulkov2022understanding}, continuous flow-matching generates ODEs with smooth, minimally varying vector fields, effectively mapping samples from a source to a target data distribution. Both diffusion and flow-matching models have delivered outstanding performance in text-to-image generation \citep{rombach2022high}, establishing new standards for creating high-quality, photorealistic images from text prompts. These models also demonstrate strong capabilities in other generative tasks, including text-to-speech, music generation, and text-to-video \citep{khachatryan2023text2video,schneider2023mo}. To accelerate sampling, techniques such as distilling pre-trained multi-step diffusion models into few-step models or developing standalone consistency models \citep{geng2025mean,liu2023instaflow,yan2024perflow,frans2024one} have been introduced, enforcing consistency constraints on network outputs across time steps to ensure stable trajectory endpoints.

\subsection{Autoregressive and Diffusion Unification}

Autoregressive and diffusion models are increasingly integrated for advanced generative modeling. Tortoise \citep{betker2023better} employs a large language model (LLM) to predict discrete, compressed mel tokens, followed by a separately trained diffusion model to upsample these coarse-grained tokens into fine-grained mel spectrograms, though this disjoint training may lead to suboptimal performance. \citep{li2024autoregressive} introduces autoregressive modeling in a continuous-valued space, defining a diffusion loss function to capture per-token probabilities from autoregressive hidden states. \citep{liu2024autoregressive} proposes encoding audio as continuous vector sequences and autoregressively generating them using a decoder-only diffusion transformer, though it operates on continuous features with large diffusion steps. \citep{yuan2024continuous} combines flow-matching loss with a pretrained autoregressive LLM and a small MLP network to predict the probability distribution of continuous-valued speech tokens from speech prompts. Lumos-1 \citep{yuan2025lumos} adopts a discrete diffusion paradigm, addressing frame-wise loss imbalances caused by spatial information redundancy in video frames. Meanwhile, the latest Gemini diffusion model \citep{team2023gemini} explores a novel language model design, enhancing user control, creativity, and efficiency in text generation.

\section{Dragon-FM}
\label{gen_inst}
The architecture of the model is depicted in Figure \ref{fig:overview}. Its inference speedup stems from three key factors: (1) the codec sequence is significantly shorter than those of prior solutions for 48 kHz audio output; (2) autoregressive iterations are substantially reduced by processing 25 tokens simultaneously, compared to single-token autoregression; and (3) flow-matching denoising steps are minimized through mean flow optimization.

\subsection{Architecture}

\paragraph{Design Principles of Fast Text-to-Speech (TTS)}
The proposed model integrates autoregressive (AR) and flow-matching paradigms into a single model, optimizing several factors for fast, high-quality text-to-speech (TTS) generation. In summary, we address two key aspects: sequence length and iteration steps.
\begin{itemize}
\item \textbf{Sequence Length:} AR and diffusion transformer (DiT) models rely on Transformer backbones, where shorter sequences reduce latency due to the quadratic complexity of self-attention. It processes a composite sequence comprising the input text, a speaker prompt, and the output speech. We design a compact discrete speech token sequence at 12.5 tokens per second, which produces high-fidelity 48 kHz audio while maintaining a short sequence length.
\item \textbf{Iteration Steps:} AR models generate tokens or blocks sequentially, while diffusion models iteratively refine tokens within a block; both contribute to latency through iterative processes. Longer token ranges increase latency, whereas shorter ranges may reduce sampling diversity and increase total processing time. To address this, the proposed model processes 2-second token blocks autoregressively, reducing AR steps compared to traditional token-by-token generation. Additionally, we employ few-step flow-matching within each block to minimize diffusion steps, enhancing parallel denoising efficiency without compromising diversity.
\end{itemize}

\paragraph{48kHz Audio Codec}

\begin{figure}
  \centering
  \includegraphics[width=1.0\textwidth]{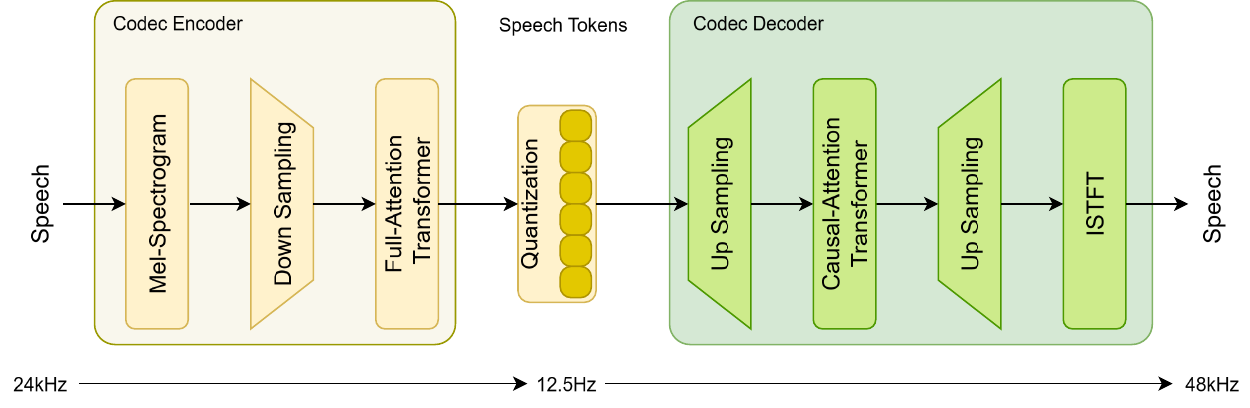}
  \caption{Overview of the asymmetric codec architecture. The codec encodes 48 kHz speech into 12.5 Hz, FSQ tokens via mel-spectrogram extraction, downsampling, and full-attention Transformers. The decoder upsamples the tokens in two stages interleaved with causal-attention Transformers and reconstructs 48 kHz audio via ISTFT.}
   \label{fig:figure3}
\end{figure}

As shown in Figure \ref{fig:figure3}, the codec model comprises four key components: a codec encoder, a finite scalar quantization (FSQ) tokenizer, a codec decoder, and two discriminators. The codec encoder first processes 48 kHz speech waveforms, extracting downsampled hidden speech representations. Then a full-attention Transformer \citep{vaswani2017attention} subsequently converts these features into 12.5 Hz latent audio representations. The FSQ tokenizer then quantizes these representations into discrete FSQ tokens. The codec decoder reconstructs the waveform in two distinct stages: first, FSQ tokens are upsampled and processed by a causal-attention Transformer; then, a second upsampling stage further increases the frame rate, followed by waveform synthesis at 48 kHz using the inverse short-time Fourier transform (ISTFT) \citep{kaneko2022istftnet}. The encoder operates bidirectionally, while the decoder is fully causal, thereby enabling streaming synthesis. The discriminator module includes two submodules: a Multi-Period Discriminator (MPD) \citep{kong2020hifi}, which captures periodic patterns across multiple temporal resolutions to enhance prosody and rhythm, and a Multi-Scale Short-Time Fourier Transform (MS-STFT) discriminator \citep{defossez2022high}, which promotes spectral fidelity and perceptual quality. Overall, the proposed codec efficiently compresses 48 kHz input speech into a compact 12.5 Hz FSQ token representation and reconstructs high-fidelity 48 kH audio.

\paragraph{Discrete Tokens as Continuous Vectors}

We avoid using continuous mel-spectrograms as bottleneck features, as empirical results indicate that mel features become unstable and difficult to reconstruct at low frame rates (e.g., 12.5 Hz). Similarly, we eschew continuous VAE-style bottlenecks in favor of finite scalar quantization (FSQ)-based discrete embeddings, which provide a structured and compact representation, significantly stabilizing and simplifying the training of the downstream denoising model.

A key insight from our work is that continuous denoising models, without requiring architectural modifications, exhibit robust intrinsic classification capabilities. By designing effective token embeddings, we leverage this latent capacity to enhance performance. This finding suggests that continuous denoising models can serve as a versatile framework for multimodal architectures, unifying discrete and continuous data representations. We view this as a promising direction for future research, particularly for large language models (LLMs) and their multimodal extensions.

\paragraph{Multi Token Autoregressive Modelling}

Autoregressive (AR) model’s latency increases significantly with longer sequence lengths due to the quadratic complexity of self-attention and additional model forward passes. Common strategies to reduce speech language model’s latency include compressing sequences into shorter token sets, which risks degrading audio quality, or using parallel codebooks and residual quantizers to enhance quality, though predicting multi-layer codes remains challenging \citep{copet2023simple}. Flattening multi-layer tokens extends sequence length, while token delays impair dependency modeling. Hierarchical modeling mitigates this by employing a first decoder to model the initial codec layer and a second decoder to recursively model subsequent layers, where the larger model typically outputs an over-compressed hidden vector for the smaller decoder. In the proposed model, we generate high-fidelity 48 kHz audio using multiple quantizers at a compact frame rate of 12.5 tokens per second, enhancing self-attention efficiency with shorter sequences. Instead of token-by-token AR prediction, we adopt chunk-wise AR modeling, where each chunk has 25 speech frames (2 seconds). Compared to EnCodec’s \citep{defossez2022high} 150 frames per 2 seconds, the attention complexity is much reduced, as is the autoregressive cost.

\paragraph{Few Step Denoising}

Flow-matching models offer a notable advantage over diffusion models by significantly reducing the number of ordinary differential equation (ODE) steps required for high-quality generation. However, autoregressive flow-matching models typically demand more iterative steps than non-autoregressive variants. To address this, we leverage mean flow \citep{geng2025mean} to drastically reduce the iteration steps per audio chunk, ensuring high generation quality with negligible degradation. While distillation techniques, such as teacher-student frameworks with consistency, present an alternative for step reduction, they incur the additional complexity and cost of training or managing a separate teacher model.

\section{Experiment}

\label{headings}

\subsection{Datasets}

To evaluate our model's performance and investigate the synergy of autoregressive and diffusion models, we curated a publicly available English podcast dataset comprising 60,000 hours of diverse audio. This dataset encompasses varied topics, speakers, and recording conditions to thoroughly assess the model's capabilities. An internal automatic speech recognition (ASR) system transcribed the audio, noisy segments were filtered, and transcriptions were standardized using text frontend rules. Audio data was segmented by speaker, and audio clips with irregular pitch or duration were excluded. Raw audio was not denoised to avoid potential signal loss from denoising models, as the speaker prompt ensures acoustic and timbre consistency between prompt and target audio.

\subsection{Training details}

To ensure maximal speaker similarity, we randomly select speaker prompts and corresponding speech targets from the same speaker and paragraph within the dataset, enabling it to capture intricate speaker-specific patterns and generate high-fidelity outputs across diverse audio samples. Training is conducted for 2 epochs, utilizing a standard exponential moving average (EMA) to stabilize model updates. We employ the Adam optimizer with an initial learning rate of 0.001, which exponentially decays to 0.0001 starting from 300,000 iterations.

\subsection{Results}

Our evaluation encompasses a codec ablation study examining varying token rates, chunk token lengths, and chunk numbers; a Fréchet Audio Distance (FAD) comparison for podcast diversity; and an inference iteration analysis across different systems.

\subsubsection{Diversity evaluation}

Podcast generation requires high diversity and natural disfluency, which traditional metrics like pitch variation or duration statistics fail to adequately capture. To assess speech diversity, we use the Fréchet Audio Distance (FAD), adapted from the Fréchet Inception Distance (FID) employed in image generation \citep{le2024voicebox}. FAD quantifies the similarity between the distributions of generated and ground-truth audio, where a smaller distance indicates a smaller gap, reflecting both higher sample quality and greater diversity. As presented in Table \ref{tab:tablediv}. Dragon-FM-A3 and Dragon-FM-A1 achieve a lower Fréchet Audio Distance (FAD) score than Dragon-FM-A2, suggesting that additional flow-matching steps enhance diversity and similarity, particularly in speech details and fidelity. Dragon-FM-B1 exhibits a lower FAD score than Dragon-FM-A1, likely due to its increased autoregressive loops, which may introduce greater sampling variation. Similarly, Dragon-FM-C1 scores lower than Dragon-FM-A1, despite using more tokens per chunk, possibly because its improved codec reconstruction contributes to a better FAD score.

\begin{table}[h]
\centering
\setlength{\tabcolsep}{8pt} 
\renewcommand{\arraystretch}{1.5} 
\begin{tabular}{ccccc} 
\toprule
\textbf{Model} & \textbf{Chunk Size} & \textbf{Token Rate} & \textbf{FM step} & \textbf{FAD} \\ \midrule
Dragon-FM-A1 & 2s & 12.5  & 12 & 2.4 \\
Dragon-FM-A2 & 2s & 12.5& 24 & 2.2   \\
Dragon-FM-A3 & 2s & 12.5& 6 & 2.6   \\
Dragon-FM-B1 & 1s & 12.5 & 12 & 2.1   \\
Dragon-FM-B2 & 1s & 12.5 & 24 & 1.9   \\
Dragon-FM-C1 & 2s & 20  & 12 & 2.2  \\
Dragon-FM-C2 & 2s & 20  & 24 & 2.0  \\ \bottomrule
\end{tabular}
\vspace{10pt}
\caption{comparison chunk size, token rate, flow matching step (FM step) and FAD.}
\label{tab:tablediv}
\end{table}

\subsubsection{RTF and latency analysis}

Autoregressive (AR) model’s inference relies on AR steps utilizing a key-value (KV) cache, whereas diffusion model inference speed depends on denoising steps without a KV cache. Given one second of audio traditional AR models like VALL-E operate on single tokens, its chunk size can be regarded as 1, while flow-matching models like E2 use a chunk size of 90. Excluding text prompt tokenization and speaker prompt size, which may influence KV cache and self-attention performance, we define acoustic model inference efficiency as the Total Number of Function Evaluations (TNFE), encompassing both AR and diffusion steps for simplicity. TNFE is the product of total AR steps and the number of diffusion iterations in each AR step.

\begin{equation} 
TNFE = Step_{\rm AR}*NFE_{\rm FM} 
\end{equation}

In this context, for two-second audio (the minimum chunk length is 2 seconds, which we use as a default for comparison with other systems), VALL-E requires 150 TNFE (for one second, 75 AR steps and 1 diffusion iteration within one AR step), and E2 requires 32 TNFE (we regard 2-second audio as one AR token for E2; E2’s flow matching step is 32), as shown in Table \ref{tab:table_ar_nfe}. Similarly, latency improvements follow this trend: VALL-E theoretically requires one AR step, while E2’s 32 TNFE introduces significantly greater latency. With 2 NFE, the proposed model achieves substantially lower latency than E2. When generating longer segments, such as 32s of audio, Dragon-FM maintains an RTF advantage over VALL-E. Although E2 and Dragon-FM may have similar RTF because their TNFEs are roughly the same, Dragon-FM's latency remains significantly lower than E2's. Cascaded systems like Tortoise require a streaming decoder for their diffusion model, necessitating complex smoothing mechanisms to prevent audio artifacts, whereas Dragon-FM’s chunk-based bidirectional attention enables chunk-wise smoothing without additional streaming design.

\begin{table}[h]
\centering
\setlength{\tabcolsep}{8pt} 
\renewcommand{\arraystretch}{1.5} 
\begin{tabular}{ccccccc} 
\toprule
\textbf{Model} & \textbf{Cache} & \textbf{TNFE} & \textbf{AR Steps} & \textbf{NFE} & \textbf{Chunk Token} & \textbf{Sampling Rate} \\ \midrule
E2 & no & 32 & 1  & 32 & 180 & 24kHz \\
VALL-E & yes & 150 & 150 & 1 & 1 & 24 kHz \\
Dragon-FM & yes & 2 & 1 & 2 & 25 & 48 kHz \\ 
Dragon-FM & yes & 4 & 1 & 4 & 25 & 48 kHz \\  \midrule 

E2 & no & 32 & 1  & 32 & 2880 & 24 kHz \\
VALL-E & yes & 2400 & 2400 & 1 & 1 & 24 kHz \\
Dragon-FM & yes & 32 & 16 & 2 & 25 & 48 kHz \\ 
Dragon-FM & yes & 64 & 16 & 4 & 25 & 48 kHz \\   \bottomrule
\end{tabular}
\vspace{10pt}
\caption{Comparison of Autoregressive and Flow-Matching Iteration Steps for a 2-Second Audio Segment. Assume the second row is 2-second audio segment to generate and the third row is 32 second audio. Dragon-FM's AR step length is 2 second audio and E2's AR step length is total audio. For VALL-E, we simplify the process by assuming each step corresponds to one diffusion step, despite VALL-E not utilizing a diffusion model, we ignore VALL-E's NAR model in this analysis.}
\label{tab:table_ar_nfe}
\end{table}

\subsubsection{Codec reconstruction comparison}

Our experimental evaluation, summarized in Table \ref{tab:table2}, evaluates the performance of our proposed Codec. Unlike traditional codecs that primarily focus on minimizing bitrate (bits/s) or token rate, our design objective is to significantly compress the frame rate while preserving high-fidelity audio reconstruction.

We established four baselines for comparison: the original recording and three versions of our internally trained Mel-spectrogram vocoder, operating at frame rates of 25 Hz, 20 Hz, and 16 Hz. The results indicate a clear trend: as the frame rate for the vocoder decreases, there is a corresponding degradation in quality, evidenced by a progressive decline in Speaker Similarity (SIM) and an increase in Word Error Rate (WER). This performance loss becomes particularly pronounced when the frame rate drops below 20 Hz (e.g., the 16 Hz model shows a SIM of 0.904 and WER of 3.01). We attribute this limitation to the short-time stationarity assumption inherent in the Short-Time Fourier Transform (STFT) process, which struggles to maintain quality at lower frame rates.

In stark contrast, our Codec demonstrates superior performance at significantly reduced frame rates. For instance, our 12.5 Hz Codec A and B maintains high reconstruction quality, outperforming the 16 Hz Mel-spectrogram baseline in terms of speaker similarity and achieving a lower WER. This highlights the advantage of our neural Codec approach, which can effectively model complex fine-grained acoustic details even at frame rates where such STFT-based methods falter.

After a comprehensive analysis of the trade-offs among reconstruction quality, the learnability of the feature distribution for a denoising model, and the degree of frame rate compression, we have selected CodecB as our default model. Operating at 12.5 Hz, it strikes an optimal balance, achieving a high SIM of 0.916 and a low WER of 2.74, thereby offering substantial frame rate reduction without a significant compromise in audio fidelity.

\begin{table}[h]
\centering
\setlength{\tabcolsep}{10pt} 
\renewcommand{\arraystretch}{1.5}
\begin{tabular}{ccccccc} 
\toprule
\textbf{Model} & \textbf{Frame Rate} & \textbf{SR} & \textbf{Bps} & \textbf{Codebook} & \textbf{SIM} & \textbf{WER} \\ 
\midrule
Recording &  &  &   &  & 1.000 & 2.23 \\ 
Mel-VocoderA & 25  & 48k &  &    & 0.937 & 2.39 \\
Mel-VocoderB & 20  & 48k &  &    & 0.926 & 2.53 \\
Mel-VocoderC & 16  & 48k &  &    & 0.904 & 3.01 \\
\midrule
CodecA & 12.5  & 48k & 3962 & 100x9  & 0.925 & 2.65 \\
CodecB & 12.5  & 48k & 2902  & 100x5  & 0.916 & 2.74 \\
CodecC & 8.3  & 48k  & 2641 & 100x9  & 0.905 & 3.14 \\
CodecD & 8.3  & 48k  & 1935 & 100x5  & 0.901 & 3.38 \\ 
\bottomrule
\end{tabular}
\vspace{8pt}
\caption{Comparison of our proposed Codec with Mel-spectrogram baselines on audio reconstruction quality. We report on Speaker Similarity (SIM) and Word Error Rate (WER) at various frame rates. Our model demonstrates superior fidelity, especially at lower frame rates. 'Recording' serves as the ground truth.}
\label{tab:table2}
\end{table}

\section{Conclusion}

In this work, we introduced Dragon-FM, a novel text-to-speech (TTS) framework that unifies autoregressive (AR) and flow-matching paradigms to address the fragmented landscape of generative modeling. By integrating a high-quality 48 kHz codec at a compact 12.5 Hz token rate, Dragon-FM achieves efficient acoustic modeling while preserving high-fidelity audio output. Our chunk-wise AR modeling, processing 2-second blocks of 25 tokens, combined with mean flow within each chunk, significantly reduces iteration steps. A key insight from Dragon-FM is the unification of discrete and continuous feature prediction, leveraging continuous vector embeddings of FSQ tokens to harness the intrinsic classification capabilities of continuous denoising models. This approach mitigates the instability of low-frame-rate mel-spectrograms and simplifies training, offering a versatile framework for multimodal generative modeling. By addressing the challenges of sequence length, iteration steps, and feature representation, Dragon-FM advances TTS and offers a framework potentially adaptable to other generative tasks. Future work will explore extending Dragon-FM’s framework to large-scale multimodal large language models (LLMs), leveraging its unified discrete-continuous approach to enable efficient, high-quality synthesis across diverse data modalities.


\bibliography{mybib}

\begin{thebibliography}{57}
\providecommand{\natexlab}[1]{#1}
\providecommand{\url}[1]{\texttt{#1}}
\expandafter\ifx\csname urlstyle\endcsname\relax
  \providecommand{\doi}[1]{doi: #1}\else
  \providecommand{\doi}{doi: \begingroup \urlstyle{rm}\Url}\fi

\bibitem[Betker(2023)]{betker2023better}
James Betker.
\newblock Better speech synthesis through scaling.
\newblock \emph{arXiv preprint arXiv:2305.07243}, 2023.

\bibitem[Brown et~al.(2020)Brown, Mann, Ryder, Subbiah, Kaplan, Dhariwal, Neelakantan, Shyam, Sastry, Askell, et~al.]{brown2020language}
Tom Brown, Benjamin Mann, Nick Ryder, Melanie Subbiah, Jared~D Kaplan, Prafulla Dhariwal, Arvind Neelakantan, Pranav Shyam, Girish Sastry, Amanda Askell, et~al.
\newblock Language models are few-shot learners.
\newblock \emph{Advances in neural information processing systems}, 33:\penalty0 1877--1901, 2020.

\bibitem[Chen et~al.(2021)Chen, Tan, Li, Liu, Qin, Zhao, and Liu]{chen2021adaspeech}
Mingjian Chen, Xu~Tan, Bohan Li, Yanqing Liu, Tao Qin, Sheng Zhao, and Tie-Yan Liu.
\newblock Adaspeech: Adaptive text to speech for custom voice.
\newblock \emph{arXiv preprint arXiv:2103.00993}, 2021.

\bibitem[Chen et~al.(2024)Chen, Liu, Zhou, Liu, Tan, Li, Zhao, Qian, and Wei]{chen2024vall}
Sanyuan Chen, Shujie Liu, Long Zhou, Yanqing Liu, Xu~Tan, Jinyu Li, Sheng Zhao, Yao Qian, and Furu Wei.
\newblock Vall-e 2: Neural codec language models are human parity zero-shot text to speech synthesizers.
\newblock \emph{arXiv preprint arXiv:2406.05370}, 2024.

\bibitem[Copet et~al.(2023)Copet, Kreuk, Gat, Remez, Kant, Synnaeve, Adi, and D{\'e}fossez]{copet2023simple}
Jade Copet, Felix Kreuk, Itai Gat, Tal Remez, David Kant, Gabriel Synnaeve, Yossi Adi, and Alexandre D{\'e}fossez.
\newblock Simple and controllable music generation.
\newblock \emph{Advances in Neural Information Processing Systems}, 36:\penalty0 47704--47720, 2023.

\bibitem[D{\'e}fossez et~al.(2022)D{\'e}fossez, Copet, Synnaeve, and Adi]{defossez2022high}
Alexandre D{\'e}fossez, Jade Copet, Gabriel Synnaeve, and Yossi Adi.
\newblock High fidelity neural audio compression.
\newblock \emph{arXiv preprint arXiv:2210.13438}, 2022.

\bibitem[Eskimez et~al.(2024)Eskimez, Wang, Thakker, Li, Tsai, Xiao, Yang, Zhu, Tang, Tan, et~al.]{eskimez2024e2}
Sefik~Emre Eskimez, Xiaofei Wang, Manthan Thakker, Canrun Li, Chung-Hsien Tsai, Zhen Xiao, Hemin Yang, Zirun Zhu, Min Tang, Xu~Tan, et~al.
\newblock E2 tts: Embarrassingly easy fully non-autoregressive zero-shot tts.
\newblock \emph{arXiv preprint arXiv:2406.18009}, 2024.

\bibitem[Frans et~al.(2024)Frans, Hafner, Levine, and Abbeel]{frans2024one}
Kevin Frans, Danijar Hafner, Sergey Levine, and Pieter Abbeel.
\newblock One step diffusion via shortcut models.
\newblock \emph{arXiv preprint arXiv:2410.12557}, 2024.

\bibitem[Furuta et~al.(2024)Furuta, Zen, Schuurmans, Faust, Matsuo, Liang, and Yang]{furuta2024improving}
Hiroki Furuta, Heiga Zen, Dale Schuurmans, Aleksandra Faust, Yutaka Matsuo, Percy Liang, and Sherry Yang.
\newblock Improving dynamic object interactions in text-to-video generation with ai feedback.
\newblock \emph{arXiv preprint arXiv:2412.02617}, 2024.

\bibitem[Geng et~al.(2025)Geng, Deng, Bai, Kolter, and He]{geng2025mean}
Zhengyang Geng, Mingyang Deng, Xingjian Bai, J~Zico Kolter, and Kaiming He.
\newblock Mean flows for one-step generative modeling.
\newblock \emph{arXiv preprint arXiv:2505.13447}, 2025.

\bibitem[Han et~al.(2024)Han, Zhou, Liu, Chen, Meng, Qian, Liu, Zhao, Li, and Wei]{han2024vall}
Bing Han, Long Zhou, Shujie Liu, Sanyuan Chen, Lingwei Meng, Yanming Qian, Yanqing Liu, Sheng Zhao, Jinyu Li, and Furu Wei.
\newblock Vall-e r: Robust and efficient zero-shot text-to-speech synthesis via monotonic alignment.
\newblock \emph{arXiv preprint arXiv:2406.07855}, 2024.

\bibitem[Jia et~al.(2025)Jia, Chen, Chen, Du, Wu, Cong, Zhuang, Li, Wei, Wang, et~al.]{jia2025ditar}
Dongya Jia, Zhuo Chen, Jiawei Chen, Chenpeng Du, Jian Wu, Jian Cong, Xiaobin Zhuang, Chumin Li, Zhen Wei, Yuping Wang, et~al.
\newblock Ditar: Diffusion transformer autoregressive modeling for speech generation.
\newblock \emph{arXiv preprint arXiv:2502.03930}, 2025.

\bibitem[Ju et~al.(2024)Ju, Wang, Shen, Tan, Xin, Yang, Liu, Leng, Song, Tang, et~al.]{ju2024naturalspeech}
Zeqian Ju, Yuancheng Wang, Kai Shen, Xu~Tan, Detai Xin, Dongchao Yang, Yanqing Liu, Yichong Leng, Kaitao Song, Siliang Tang, et~al.
\newblock Naturalspeech 3: Zero-shot speech synthesis with factorized codec and diffusion models.
\newblock \emph{arXiv preprint arXiv:2403.03100}, 2024.

\bibitem[Kanda et~al.(2024)Kanda, Wang, Eskimez, Thakker, Yang, Zhu, Tang, Li, Tsai, Xiao, et~al.]{kanda2024making}
Naoyuki Kanda, Xiaofei Wang, Sefik~Emre Eskimez, Manthan Thakker, Hemin Yang, Zirun Zhu, Min Tang, Canrun Li, Chung-Hsien Tsai, Zhen Xiao, et~al.
\newblock Making flow-matching-based zero-shot text-to-speech laugh as you like.
\newblock \emph{arXiv preprint arXiv:2402.07383}, 2024.

\bibitem[Kaneko et~al.(2022)Kaneko, Tanaka, Kameoka, and Seki]{kaneko2022istftnet}
Takuhiro Kaneko, Kou Tanaka, Hirokazu Kameoka, and Shogo Seki.
\newblock istftnet: Fast and lightweight mel-spectrogram vocoder incorporating inverse short-time fourier transform.
\newblock In \emph{ICASSP 2022-2022 IEEE International Conference on Acoustics, Speech and Signal Processing (ICASSP)}, pages 6207--6211. IEEE, 2022.

\bibitem[Khachatryan et~al.(2023)Khachatryan, Movsisyan, Tadevosyan, Henschel, Wang, Navasardyan, and Shi]{khachatryan2023text2video}
Levon Khachatryan, Andranik Movsisyan, Vahram Tadevosyan, Roberto Henschel, Zhangyang Wang, Shant Navasardyan, and Humphrey Shi.
\newblock Text2video-zero: Text-to-image diffusion models are zero-shot video generators.
\newblock In \emph{Proceedings of the IEEE/CVF International Conference on Computer Vision}, pages 15954--15964, 2023.

\bibitem[Khrulkov et~al.(2022)Khrulkov, Ryzhakov, Chertkov, and Oseledets]{khrulkov2022understanding}
Valentin Khrulkov, Gleb Ryzhakov, Andrei Chertkov, and Ivan Oseledets.
\newblock Understanding ddpm latent codes through optimal transport.
\newblock \emph{arXiv preprint arXiv:2202.07477}, 2022.

\bibitem[Kim et~al.(2021)Kim, Kong, and Son]{kim2021conditional}
Jaehyeon Kim, Jungil Kong, and Juhee Son.
\newblock Conditional variational autoencoder with adversarial learning for end-to-end text-to-speech.
\newblock In \emph{International Conference on Machine Learning}, pages 5530--5540. PMLR, 2021.

\bibitem[Kong et~al.(2020)Kong, Kim, and Bae]{kong2020hifi}
Jungil Kong, Jaehyeon Kim, and Jaekyoung Bae.
\newblock Hifi-gan: Generative adversarial networks for efficient and high fidelity speech synthesis.
\newblock \emph{Advances in neural information processing systems}, 33:\penalty0 17022--17033, 2020.

\bibitem[Le et~al.(2024)Le, Vyas, Shi, Karrer, Sari, Moritz, Williamson, Manohar, Adi, Mahadeokar, et~al.]{le2024voicebox}
Matthew Le, Apoorv Vyas, Bowen Shi, Brian Karrer, Leda Sari, Rashel Moritz, Mary Williamson, Vimal Manohar, Yossi Adi, Jay Mahadeokar, et~al.
\newblock Voicebox: Text-guided multilingual universal speech generation at scale.
\newblock \emph{Advances in neural information processing systems}, 36, 2024.

\bibitem[Leng et~al.(2023)Leng, Guo, Shen, Tan, Ju, Liu, Liu, Yang, Zhang, Song, et~al.]{leng2023prompttts}
Yichong Leng, Zhifang Guo, Kai Shen, Xu~Tan, Zeqian Ju, Yanqing Liu, Yufei Liu, Dongchao Yang, Leying Zhang, Kaitao Song, et~al.
\newblock Prompttts 2: Describing and generating voices with text prompt.
\newblock \emph{arXiv preprint arXiv:2309.02285}, 2023.

\bibitem[Li et~al.(2024{\natexlab{a}})Li, Wang, Wang, Qian, Zhou, Liu, Yousefi, Li, Tsai, Xiao, et~al.]{li2024investigating}
Jiaqi Li, Dongmei Wang, Xiaofei Wang, Yao Qian, Long Zhou, Shujie Liu, Midia Yousefi, Canrun Li, Chung-Hsien Tsai, Zhen Xiao, et~al.
\newblock Investigating neural audio codecs for speech language model-based speech generation.
\newblock \emph{arXiv preprint arXiv:2409.04016}, 2024{\natexlab{a}}.

\bibitem[Li et~al.(2018)Li, Liu, Liu, Zhao, Liu, and Zhou]{li2018close}
Naihan Li, Shujie Liu, Yanqing Liu, Sheng Zhao, Ming Liu, and Ming Zhou.
\newblock Close to human quality tts with transformer.
\newblock \emph{arXiv preprint arXiv:1809.08895}, 2, 2018.

\bibitem[Li et~al.(2019)Li, Liu, Liu, Zhao, and Liu]{li2019neural}
Naihan Li, Shujie Liu, Yanqing Liu, Sheng Zhao, and Ming Liu.
\newblock Neural speech synthesis with transformer network.
\newblock In \emph{Proceedings of the AAAI conference on artificial intelligence}, volume~33, pages 6706--6713, 2019.

\bibitem[Li et~al.(2020{\natexlab{a}})Li, Liu, Liu, Zhao, Liu, and Zhou]{li2020moboaligner}
Naihan Li, Shujie Liu, Yanqing Liu, Sheng Zhao, Ming Liu, and Ming Zhou.
\newblock Moboaligner: A neural alignment model for non-autoregressive tts with monotonic boundary search.
\newblock \emph{arXiv preprint arXiv:2005.08528}, 2020{\natexlab{a}}.

\bibitem[Li et~al.(2020{\natexlab{b}})Li, Liu, Wu, Liu, Zhao, and Liu]{li2020robutrans}
Naihan Li, Yanqing Liu, Yu~Wu, Shujie Liu, Sheng Zhao, and Ming Liu.
\newblock Robutrans: A robust transformer-based text-to-speech model.
\newblock In \emph{Proceedings of the AAAI conference on artificial intelligence}, volume~34, pages 8228--8235, 2020{\natexlab{b}}.

\bibitem[Li et~al.(2024{\natexlab{b}})Li, Tian, Li, Deng, and He]{li2024autoregressive}
Tianhong Li, Yonglong Tian, He~Li, Mingyang Deng, and Kaiming He.
\newblock Autoregressive image generation without vector quantization.
\newblock \emph{arXiv preprint arXiv:2406.11838}, 2024{\natexlab{b}}.

\bibitem[Lipman et~al.(2022)Lipman, Chen, Ben-Hamu, Nickel, and Le]{lipman2022flow}
Yaron Lipman, Ricky~TQ Chen, Heli Ben-Hamu, Maximilian Nickel, and Matt Le.
\newblock Flow matching for generative modeling.
\newblock \emph{arXiv preprint arXiv:2210.02747}, 2022.

\bibitem[Liu et~al.(2022{\natexlab{a}})Liu, Gong, and Liu]{liu2022flow}
Xingchao Liu, Chengyue Gong, and Qiang Liu.
\newblock Flow straight and fast: Learning to generate and transfer data with rectified flow.
\newblock \emph{arXiv preprint arXiv:2209.03003}, 2022{\natexlab{a}}.

\bibitem[Liu et~al.(2023)Liu, Zhang, Ma, Peng, et~al.]{liu2023instaflow}
Xingchao Liu, Xiwen Zhang, Jianzhu Ma, Jian Peng, et~al.
\newblock Instaflow: One step is enough for high-quality diffusion-based text-to-image generation.
\newblock In \emph{The Twelfth International Conference on Learning Representations}, 2023.

\bibitem[Liu et~al.(2021)Liu, Xu, Wang, Chen, Li, Tan, et~al.]{liu2021delightfultts}
Y~Liu, Z~Xu, G~Wang, K~Chen, B~Li, X~Tan, et~al.
\newblock Delightfultts: The microsoft speech synthesis system for blizzard challenge.
\newblock \emph{DelightfulTTS: The Microsoft Speech Synthesis System for Blizzard Challenge}, 2021.

\bibitem[Liu et~al.(2022{\natexlab{b}})Liu, Xue, He, Tan, and Zhao]{liu2022delightfultts}
Yanqing Liu, Ruiqing Xue, Lei He, Xu~Tan, and Sheng Zhao.
\newblock Delightfultts 2: End-to-end speech synthesis with adversarial vector-quantized auto-encoders.
\newblock \emph{arXiv preprint arXiv:2207.04646}, 2022{\natexlab{b}}.

\bibitem[Liu et~al.(2024)Liu, Wang, Inoue, Bai, and Li]{liu2024autoregressive}
Zhijun Liu, Shuai Wang, Sho Inoue, Qibing Bai, and Haizhou Li.
\newblock Autoregressive diffusion transformer for text-to-speech synthesis.
\newblock \emph{arXiv preprint arXiv:2406.05551}, 2024.

\bibitem[Meng et~al.(2024)Meng, Zhou, Liu, Chen, Han, Hu, Liu, Li, Zhao, Wu, et~al.]{meng2024autoregressive}
Lingwei Meng, Long Zhou, Shujie Liu, Sanyuan Chen, Bing Han, Shujie Hu, Yanqing Liu, Jinyu Li, Sheng Zhao, Xixin Wu, et~al.
\newblock Autoregressive speech synthesis without vector quantization.
\newblock \emph{arXiv preprint arXiv:2407.08551}, 2024.

\bibitem[Mentzer et~al.(2023)Mentzer, Minnen, Agustsson, and Tschannen]{mentzer2023finite}
Fabian Mentzer, David Minnen, Eirikur Agustsson, and Michael Tschannen.
\newblock Finite scalar quantization: Vq-vae made simple.
\newblock \emph{arXiv preprint arXiv:2309.15505}, 2023.

\bibitem[Parker et~al.(2024)Parker, Smirnov, Pons, Carr, Zukowski, Evans, and Liu]{parker2024scaling}
Julian~D Parker, Anton Smirnov, Jordi Pons, CJ~Carr, Zack Zukowski, Zach Evans, and Xubo Liu.
\newblock Scaling transformers for low-bitrate high-quality speech coding.
\newblock \emph{arXiv preprint arXiv:2411.19842}, 2024.

\bibitem[Radford et~al.(2018)Radford, Narasimhan, Salimans, Sutskever, et~al.]{radford2018improving}
Alec Radford, Karthik Narasimhan, Tim Salimans, Ilya Sutskever, et~al.
\newblock Improving language understanding by generative pre-training.
\newblock 2018.

\bibitem[Ramesh et~al.(2021)Ramesh, Pavlov, Goh, Gray, Voss, Radford, Chen, and Sutskever]{ramesh2021zero}
Aditya Ramesh, Mikhail Pavlov, Gabriel Goh, Scott Gray, Chelsea Voss, Alec Radford, Mark Chen, and Ilya Sutskever.
\newblock Zero-shot text-to-image generation.
\newblock In \emph{International conference on machine learning}, pages 8821--8831. Pmlr, 2021.

\bibitem[Rombach et~al.(2022)Rombach, Blattmann, Lorenz, Esser, and Ommer]{rombach2022high}
Robin Rombach, Andreas Blattmann, Dominik Lorenz, Patrick Esser, and Bj{\"o}rn Ommer.
\newblock High-resolution image synthesis with latent diffusion models.
\newblock In \emph{Proceedings of the IEEE/CVF conference on computer vision and pattern recognition}, pages 10684--10695, 2022.

\bibitem[Schneider et~al.(2023)Schneider, Kamal, Jin, and Sch{\"o}lkopf]{schneider2023mo}
Flavio Schneider, Ojasv Kamal, Zhijing Jin, and Bernhard Sch{\"o}lkopf.
\newblock Mo$\backslash$\^{} usai: Text-to-music generation with long-context latent diffusion.
\newblock \emph{arXiv preprint arXiv:2301.11757}, 2023.

\bibitem[Shen et~al.(2023)Shen, Ju, Tan, Liu, Leng, He, Qin, Zhao, and Bian]{shen2023naturalspeech}
Kai Shen, Zeqian Ju, Xu~Tan, Yanqing Liu, Yichong Leng, Lei He, Tao Qin, Sheng Zhao, and Jiang Bian.
\newblock Naturalspeech 2: Latent diffusion models are natural and zero-shot speech and singing synthesizers.
\newblock \emph{arXiv preprint arXiv:2304.09116}, 2023.

\bibitem[Sun et~al.(2025)Sun, Hu, Liu, Meng, Wang, Han, Yang, Liu, Zhao, Lu, et~al.]{sun2025zero}
Haiyang Sun, Shujie Hu, Shujie Liu, Lingwei Meng, Hui Wang, Bing Han, Yifan Yang, Yanqing Liu, Sheng Zhao, Yan Lu, et~al.
\newblock Zero-shot streaming text to speech synthesis with transducer and auto-regressive modeling.
\newblock \emph{arXiv preprint arXiv:2505.19669}, 2025.

\bibitem[Tan et~al.(2024)Tan, Chen, Liu, Cong, Zhang, Liu, Wang, Leng, Yi, He, et~al.]{tan2024naturalspeech}
Xu~Tan, Jiawei Chen, Haohe Liu, Jian Cong, Chen Zhang, Yanqing Liu, Xi~Wang, Yichong Leng, Yuanhao Yi, Lei He, et~al.
\newblock Naturalspeech: End-to-end text-to-speech synthesis with human-level quality.
\newblock \emph{IEEE Transactions on Pattern Analysis and Machine Intelligence}, 2024.

\bibitem[Team et~al.(2023)Team, Anil, Borgeaud, Alayrac, Yu, Soricut, Schalkwyk, Dai, Hauth, Millican, et~al.]{team2023gemini}
Gemini Team, Rohan Anil, Sebastian Borgeaud, Jean-Baptiste Alayrac, Jiahui Yu, Radu Soricut, Johan Schalkwyk, Andrew~M Dai, Anja Hauth, Katie Millican, et~al.
\newblock Gemini: a family of highly capable multimodal models.
\newblock \emph{arXiv preprint arXiv:2312.11805}, 2023.

\bibitem[Tong et~al.(2023)Tong, Fatras, Malkin, Huguet, Zhang, Rector-Brooks, Wolf, and Bengio]{tong2023improving}
Alexander Tong, Kilian Fatras, Nikolay Malkin, Guillaume Huguet, Yanlei Zhang, Jarrid Rector-Brooks, Guy Wolf, and Yoshua Bengio.
\newblock Improving and generalizing flow-based generative models with minibatch optimal transport.
\newblock \emph{arXiv preprint arXiv:2302.00482}, 2023.

\bibitem[Vaswani et~al.(2017)Vaswani, Shazeer, Parmar, Uszkoreit, Jones, Gomez, Kaiser, and Polosukhin]{vaswani2017attention}
Ashish Vaswani, Noam Shazeer, Niki Parmar, Jakob Uszkoreit, Llion Jones, Aidan~N Gomez, {\L}ukasz Kaiser, and Illia Polosukhin.
\newblock Attention is all you need.
\newblock \emph{Advances in neural information processing systems}, 30, 2017.

\bibitem[Wang et~al.(2023)Wang, Chen, Wu, Zhang, Zhou, Liu, Chen, Liu, Wang, Li, et~al.]{wang2023neural}
Chengyi Wang, Sanyuan Chen, Yu~Wu, Ziqiang Zhang, Long Zhou, Shujie Liu, Zhuo Chen, Yanqing Liu, Huaming Wang, Jinyu Li, et~al.
\newblock Neural codec language models are zero-shot text to speech synthesizers.
\newblock \emph{arXiv preprint arXiv:2301.02111}, 2023.

\bibitem[Wang et~al.(2025{\natexlab{a}})Wang, Liu, Meng, Li, Yang, Zhao, Sun, Liu, Sun, Zhou, et~al.]{wang2025felle}
Hui Wang, Shujie Liu, Lingwei Meng, Jinyu Li, Yifan Yang, Shiwan Zhao, Haiyang Sun, Yanqing Liu, Haoqin Sun, Jiaming Zhou, et~al.
\newblock Felle: Autoregressive speech synthesis with token-wise coarse-to-fine flow matching.
\newblock \emph{arXiv preprint arXiv:2502.11128}, 2025{\natexlab{a}}.

\bibitem[Wang et~al.(2025{\natexlab{b}})Wang, Yang, Liu, Li, Meng, Liu, Zhou, Sun, Lu, and Qin]{wang2025streammel}
Hui Wang, Yifan Yang, Shujie Liu, Jinyu Li, Lingwei Meng, Yanqing Liu, Jiaming Zhou, Haoqin Sun, Yan Lu, and Yong Qin.
\newblock Streammel: Real-time zero-shot text-to-speech via interleaved continuous autoregressive modeling.
\newblock \emph{arXiv preprint arXiv:2506.12570}, 2025{\natexlab{b}}.

\bibitem[Xue et~al.(2023)Xue, Liu, He, Tan, Liu, Lin, and Zhao]{xue2023foundationtts}
Ruiqing Xue, Yanqing Liu, Lei He, Xu~Tan, Linquan Liu, Edward Lin, and Sheng Zhao.
\newblock Foundationtts: Text-to-speech for asr customization with generative language model.
\newblock \emph{arXiv preprint arXiv:2303.02939}, 2023.

\bibitem[Yan et~al.(2024)Yan, Liu, Pan, Liew, Liu, and Feng]{yan2024perflow}
Hanshu Yan, Xingchao Liu, Jiachun Pan, Jun~Hao Liew, Qiang Liu, and Jiashi Feng.
\newblock Perflow: Piecewise rectified flow as universal plug-and-play accelerator.
\newblock \emph{Advances in Neural Information Processing Systems}, 37:\penalty0 78630--78652, 2024.

\bibitem[Yang et~al.(2025)Yang, Liu, Li, Hu, Wu, Wang, Yu, Meng, Sun, Liu, et~al.]{yang2025pseudo}
Yifan Yang, Shujie Liu, Jinyu Li, Yuxuan Hu, Haibin Wu, Hui Wang, Jianwei Yu, Lingwei Meng, Haiyang Sun, Yanqing Liu, et~al.
\newblock Pseudo-autoregressive neural codec language models for efficient zero-shot text-to-speech synthesis.
\newblock \emph{arXiv preprint arXiv:2504.10352}, 2025.

\bibitem[Yuan et~al.(2025)Yuan, Chen, Cen, Yu, Liang, Chang, Lin, Feng, Liu, Xing, et~al.]{yuan2025lumos}
Hangjie Yuan, Weihua Chen, Jun Cen, Hu~Yu, Jingyun Liang, Shuning Chang, Zhihui Lin, Tao Feng, Pengwei Liu, Jiazheng Xing, et~al.
\newblock Lumos-1: On autoregressive video generation from a unified model perspective.
\newblock \emph{arXiv preprint arXiv:2507.08801}, 2025.

\bibitem[Yuan et~al.(2024)Yuan, Liu, Liu, and Zhao]{yuan2024continuous}
Ze~Yuan, Yanqing Liu, Shujie Liu, and Sheng Zhao.
\newblock Continuous speech tokens makes llms robust multi-modality learners.
\newblock \emph{arXiv preprint arXiv:2412.04917}, 2024.

\bibitem[Zhang et~al.(2024)Zhang, Liu, Zheng, and Zhao]{zhang2024boosting}
Chong Zhang, Yanqing Liu, Yang Zheng, and Sheng Zhao.
\newblock Boosting diffusion model for spectrogram up-sampling in text-to-speech: An empirical study.
\newblock \emph{arXiv preprint arXiv:2406.04633}, 2024.

\bibitem[Zhang et~al.(2022)Zhang, Song, Tan, Tan, Yan, Liu, Wang, Zhou, Qin, Lee, et~al.]{zhang2022mixed}
Guangyan Zhang, Kaitao Song, Xu~Tan, Daxin Tan, Yuzi Yan, Yanqing Liu, Gang Wang, Wei Zhou, Tao Qin, Tan Lee, et~al.
\newblock Mixed-phoneme bert: Improving bert with mixed phoneme and sup-phoneme representations for text to speech.
\newblock \emph{arXiv preprint arXiv:2203.17190}, 2022.

\bibitem[Zhang et~al.(2023)Zhang, Zhou, Wang, Chen, Wu, Liu, Chen, Liu, Wang, Li, et~al.]{zhang2023speak}
Ziqiang Zhang, Long Zhou, Chengyi Wang, Sanyuan Chen, Yu~Wu, Shujie Liu, Zhuo Chen, Yanqing Liu, Huaming Wang, Jinyu Li, et~al.
\newblock Speak foreign languages with your own voice: Cross-lingual neural codec language modeling.
\newblock \emph{arXiv preprint arXiv:2303.03926}, 2023.

\end{thebibliography}
\bibliographystyle{plainnat}



\end{document}